\documentclass[journal]{IEEEtran}
\usepackage{hyperref}
\usepackage{graphicx}
\ifCLASSINFOpdf
\else
\fi
\usepackage{amsfonts}
\usepackage{color}
\usepackage[dvipsnames]{xcolor}
\usepackage{soul}
\usepackage[noadjust]{cite}

\begin{document}
\title{A Flexible Machine Learning-Aware Architecture\\ for Future WLANs}

\author{Francesc~Wilhelmi,~Sergio~Barrachina-Mu\~noz,~Boris~Bellalta,~Cristina~Cano,~Anders~Jonsson,~and~Vishnu~Ram
\thanks{Francesc Wilhelmi, Sergio Barrachina-Mu\~noz, Boris Bellalta, and Anders Jonsson are with Universitat Pompeu Fabra (UPF); Cristina Cano is with Universitat Oberta de Catalunya (UOC); Vishnu Ram is currently working as an independent researcher.}
}

\maketitle

\begin{abstract}
Lots of hopes have been placed on Machine Learning (ML) as a key enabler of future wireless networks. By taking advantage of large volumes of data, ML is expected to deal with the ever-increasing complexity of networking problems. Unfortunately, current networks are not yet prepared to support the ensuing requirements of ML-based applications in terms of data collection, processing, and output distribution. This article points out the architectural requirements that are needed to pervasively include ML as part of future wireless networks operation. Specifically, we look into Wireless Local Area Networks (WLANs), which, due to their nature can be found in multiple forms, ranging from cloud-based to edge-computing-like deployments. In particular, we propose to adopt the International Telecommunications Union (ITU) unified architecture for 5G and beyond. Based on ITU's architecture, we provide insights on the main requirements and the major challenges of introducing ML to the multiple modalities of WLANs. Finally, we showcase the superiority of the architecture through an ML-enabled use case for future networks.
\end{abstract}


\IEEEpeerreviewmaketitle

\begin{table*}[t!]
	\caption{Machine learning methods, algorithms, potential networking applications, and examples of input data.}
	\label{table:ml_taxonomy}
	\centering
	\begin{tabular}{|p{.09\textwidth}|p{.28\textwidth}|p{.25\textwidth}|p{.25\textwidth}|}
		\hline
		\textbf{ML method} & \textbf{Algorithms} & \textbf{Potential applications} & \textbf{Examples of input data} \\\hline
		Supervised learning & Linear classifiers, regression methods such as Autoregressive Integrated Moving Average (ARIMA), neural networks, hidden Markov models, random forest, support vector machines, k-nearest neighbors, principal component analysis &Traffic forecasting, mobility pattern prediction, flow classification, routing, anomaly detection, spectrum management, failure detection, QoE prediction & IP traffic matrices, temporal user location, availability of routing paths, application data, channel measurements, performance metrics \\\hline
		Unsupervised learning & Clustering, mixture models, generative models, non-negative matrix factorization, evolutionary algorithms & Traffic classification, virtual topology design, path computation, intruder detection, signal separation& IP traffic matrices, historical end-to-end bit-rate, received symbols \\\hline
		Reinforcement learning & Monte Carlo, Q-learning, State-Action-Reward-State-Action (SARSA), deep Q network, actor-critic, multi-armed bandits, learning automaton, Markov decision processes & Power control, rate adaptation, routing, channel selection, spatial reuse, smart caching, traffic offloading, cognitive channel access, energy harvesting, energy efficiency & Channel measurements, link status, performance metrics (e.g., throughput, delay), server occupation, power consumption \\\hline
	\end{tabular}
\end{table*}

\section{Introduction}
Wireless communications have reached a point where a paradigm shift is required to satisfy the increasing needs of next-generation applications \cite{osseiran2014scenarios}. Based on the current trend, Artificial Intelligence (AI), and more precisely Machine Learning (ML), is expected to conduct a revolution, especially regarding the network planning, operation, and management of the 5th and 6th generations (5G/6G) of mobile communications. 

ML is meant to empower a computational system for learning automatically, based on experience, so that future situations can be properly managed without having been programmed explicitly. Concerning wireless communications, there is a huge amount of unexploited data generated at both infrastructure and user levels, which could be extremely useful for learning complex patterns, thus improving network performance. For instance, the behavior of users in a network-oriented service can be predicted through ML given the information from previous sessions. Based on these predictions, network resources can be appropriately accommodated in future sessions.

Unfortunately, the potential benefits of ML for real networks are currently limited by the existing infrastructure, which is not yet prepared to accommodate ML-oriented tasks such as data collection, processing, and output distribution. Instead, current networking systems are typically meant for delivering content, without taking into account the underlying characteristics of the processes that generate it.

The first steps towards AI-enabled networking are currently being made in 5G through Network Function Virtualization (NFV). Unlike traditional hardware-based networks, NFV allows rapid elasticity and fast reconfiguration on assigning network resources. This is particularly useful to enable verticals such as autonomous driving in the automotive sector or smart manufacturing in Industry 4.0. Besides, network virtualization is useful to boost inter-operator coordination and bringing the ML operation to a macro-scale level, counting with vast information and computation resources. 

To conduct the evolution towards ML-aware networks, standardization is key to reach consensus between operators and manufacturers. In this regard, we find many initiatives held by standardization organizations, from which we highlight the Focus Group on Machine Learning for Future Networks including 5G (FG-ML5G), which belongs to the International Telecommunication Union Telecommunication Standardization Sector (ITU-T). The FG-ML5G aims to enable the convergence of future communications with ML technologies. To that end, the focus group has released a specification on a \emph{Unified architecture for 5G and beyond}, recently turned into an ITU Recommendation \cite{itu2019architecture}. Remarkably, ITU's standardized architecture provides a common nomenclature for ML-related mechanisms so that interoperability with other networking systems is achieved. 

Apart from the ITU-T initiatives, other important standardization bodies such as the 3rd Generation Partnership Project (3GPP) or the European Telecommunications Standards Institute (ETSI) are currently working on the integration of data analytics to network functions. The 3GPP contemplates AI as one of the priority topics for shaping its upcoming release (Release 17) and architectural requirements are currently under study \cite{3gpp2019study}. Furthermore, we highlight the ETSI groups on Experiential Networked Intelligence (ENI) and Zero-touch network and Service Management (ZSM), which actively study the integration of AI to networks \cite{etsi2019architecture}. Unlike the ITU's unified architecture, most of the work held by the 3GPP and the ETSI focuses on centralized data collection and data analytics solutions. Nevertheless, we understand that the works in \cite{itu2019architecture, 3gpp2019study, etsi2019architecture} are complementary to each other.

To contribute to the evolution of wireless communications towards AI-based systems, we provide a realization of the ITU's architecture for IEEE 802.11 Wireless Local Area Networks (WLANs), which will be an essential part of the 5G/6G ecosystem in the unlicensed bands. Unlike for cellular networks, WLANs have received much less attention when designing AI-aware architectural solutions. The fact is that cellular-based deployments fit in perfectly with big data analytics, due to the vast amount of data and high computation resources available for mobile network operators. In contrast, WLANs pose a set of specific challenges resulting from their multiple deployment modes (e.g., campus network, residential usage) and their typical decentralized nature. Despite WLANs can count with plenty of data to be used by ML methods in large and planned deployments, we find other residential-type scenarios that lack of powerful centralized equipment. In these cases, huge computing and processing resources cannot be endowed to the ML operation. 

To enable the integration of ML-based approaches into the different modalities of WLANs, the module-based ITU's architecture allows adapting to the problem instance and the set of available resources, thus providing flexibility in terms of deployment heterogeneity. For instance, despite deep learning is a powerful solution that may improve the performance in multiple scenarios, it entails a set of computation, storage and communication requirements that may not be fulfilled in other deployments, or parts of the network.

The main contributions of this paper are as follows:
\begin{itemize}
	\item We devise and discuss the potential of ML-enabled future communications. Then, we focus on IEEE 802.11 WLANs and provide a set of use cases. 
	\item We provide an overview of the ITU's ML-aware architecture for 5G networks and beyond.
	\item We adopt the module-based ITU's architecture and provide a realization for IEEE 802.11 WLANs, thus pointing out the major technical challenges and opportunities.
	\item We depict the potential advantages of ML-based approaches enabled by the architecture through numerical results in a particular use case.	
\end{itemize}

\section{Machine Learning as Enabler of Future Wireless Networks} 
\label{section:intro_ML}
In this section, we discuss the role of ML for sustaining the progress of future wireless networks. Then, we motivate the application of ML to IEEE 802.11 WLANs through a set of illustrative ML-driven use cases.

\subsection{Machine Learning in Communications}
The proliferation of new communication-based applications is defining the shape of future networks through a set of strict requirements \cite{itu2019use}. Some examples are Vehicle to Everything (V2X), Industry 4.0, and Virtual Reality / Augmented Reality (VR/AR). These applications are really challenging in terms of bandwidth (10-20 Gbps), latency ($<$5 ms), reliability (1 packet lost for every 10$^5$ packets sent), and scalability (1,000,000 devices/km$^2$), as well as require a flexible network response to cope with the high heterogeneity of devices and contents.

In 5G, the previous concepts are referred to as Enhanced Mobile Broadband (eMBB), Massive Machine to Machine and Internet of Things (IoT) Communication (mMTC), and Ultra-Reliable and Low Latency Communication (uRLLC), respectively. Similarly, IEEE 802.11 groups are also considering these aspects in the design of next-generation amendments, such as High Efficiency (HE) IEEE 802.11ax and Extreme High Throughput (EHT) IEEE 802.11be.

To meet the abovementioned strict requirements, not only a technological innovation is required (e.g., use of more spectrum or improve multiple-antennas technologies), but a paradigm shift is necessary when designing novel solutions for network planning, operation, and management. In particular, intelligent wireless networks need to be empowered with cognitive and context-aware capabilities, which may require additional infrastructure such as environmental sensors and cameras. To that end, ML is expected to be important during the lifetime of 5G and will become pervasive - as included from the beginning in their conception - in 6G networks. 

The actual utility of ML lies in those problems that are hard to solve by hand-programming due to their underlying complex patterns (e.g., network traffic prediction). Formally, a machine is said to learn if it improves the performance $\mathbb{P}$ obtained from undertaking task $\mathbb{T}$, based on the gathered experience $\mathbb{E}$ \cite{mitchell1997machine}. Different ML techniques have been categorized in multiple ways, but the most common taxonomy differentiates between supervised learning (labeled data is used for training), unsupervised learning (no labels are used on input data), and reinforcement learning (exploration-exploitation trade-off with label/unlabeled data). Table \ref{table:ml_taxonomy} provides a list of algorithms and potential networking applications for each type of ML techniques, as well as some examples of input data to be used by these methods. For further details, we address the interested reader to \cite{jiang2016machine, zhang2019deep, usama2019unsupervised}, which survey a plethora of ML-based applications for networking.

Apart from the specific ML solutions to problems in communications, some efforts have been made towards enabling AI-aware networking in more general terms. In particular, several architectural proposals have been provided so far \cite{bi2015wireless,chih2017big,wang2018machine}. Most of the referenced works agree in the necessary steps for enabling big data analytics in cellular deployments: (1) data collection, (2) data preparation, (3) data analysis, and (4) decision making. Nevertheless, none of these works provide architectural guidelines to introduce ML to wireless networks. In this regard, the ITU's architecture looks deeper into the ML operation and targets the actual procedures involving information gathering, processing, and communication. Besides, the ITU-T provides a data handling framework for ML-aware networks \cite{itu2019data}, which defines processes concerning data collection, processing, and output distribution. 

\subsection{Machine Learning-Enabled Use Cases in WLANs}
To showcase the potential of applying AI in IEEE 802.11 WLANs, we next describe a set of use cases where ML allows improving the network operation.

\subsubsection{OFDMA-Based Smart Network Slicing} 
Network slicing is one of the hottest topics in 5G because it allows virtually separating network resources to meet diverse application requirements. In next-generation WLANs, network slicing can be realized through the allocation of radio resources via Orthogonal Frequency-Division Multiple Access (OFDMA). However, the heterogeneity of applications and devices, and their subsequent elasticity prevent allocating frequency resources easily. To solve this, ML can be used to make predictions on the user requirements so that the access network can be optimized.

\begin{figure}[ht!]
	\centering
	\includegraphics[width=1\columnwidth]{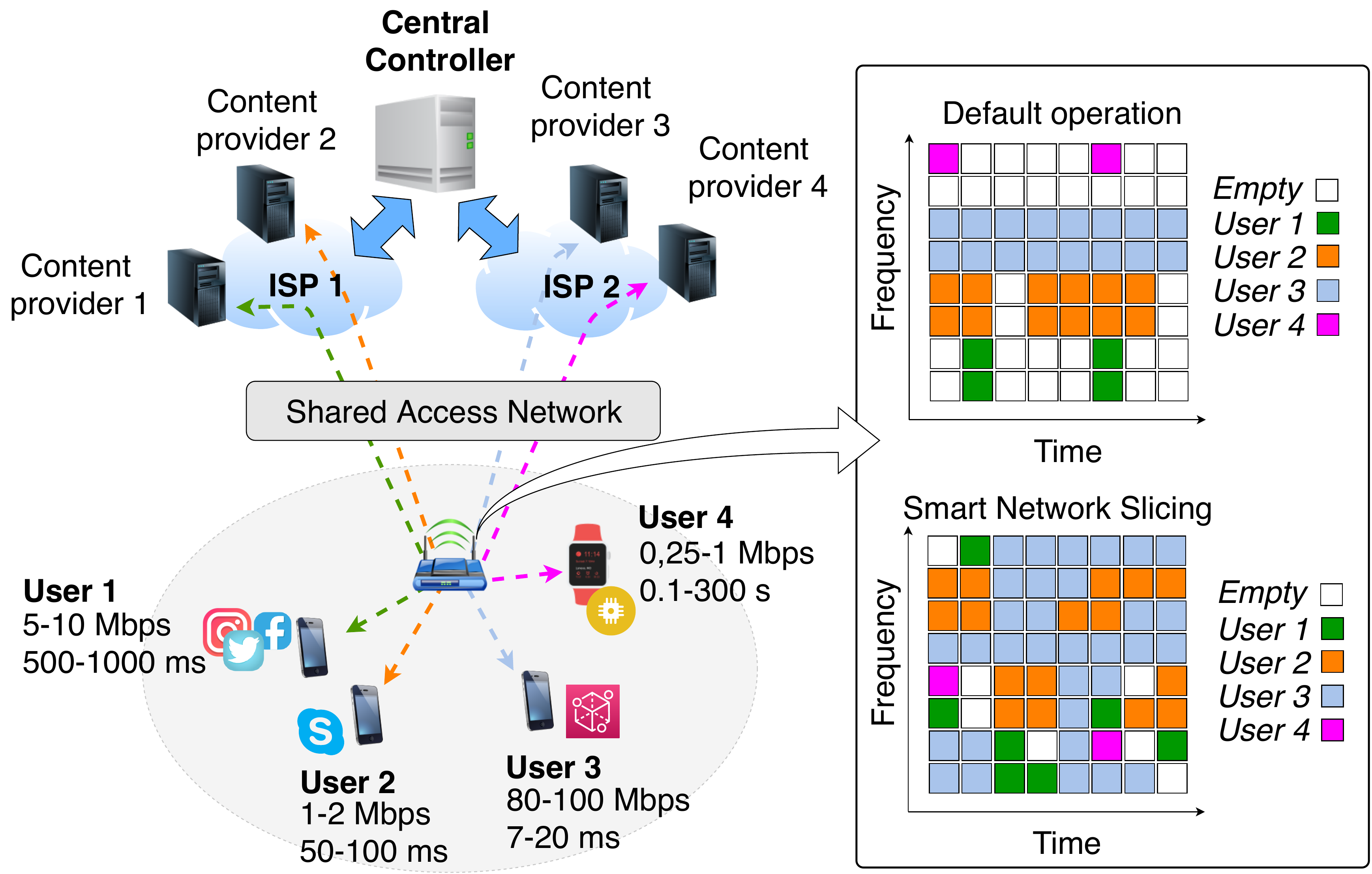}
	\caption{OFDMA-based smart network slicing.}
	\label{fig:use_cases}
\end{figure}

As an example, Fig. \ref{fig:use_cases} shows a scenario where multiple users operate under different requirements, based on the applications they use. While the central controller can make predictions on user behavior, the local schedulers may consider information such as the user profile, the current performance, and the environmental circumstances. Accordingly, the Access Point (AP) can allocate the most suitable OFDMA resources to each device, based on the predicted needs and network status.

\subsubsection{Cloud-Based User Association and Handover}
Most of the current user association and handover procedures held in WLANs typically rely on the Strongest Signal First (SSF) mechanism. This might be problematic in terms of load balancing and can potentially lead to severe performance degradation in dense Basic Service Sets (BSSs). By introducing ML, it is possible to handle contextual information such as the traffic load, which can be useful for decision-making. Furthermore, mobility pattern prediction and user requirements forecasts can be included in the system, thus empowering the association and handover mechanisms with insightful information.

\subsubsection{Inference-Based Coordinated Scheduling}
Contrary to traditional cellular-type networks, WLAN deployments can be chaotic, especially in residential scenarios where anyone can set-up an AP and create a wireless network. This typically leads to complex scenarios where inter-BSS interactions prevent the existing scheduling approaches to ensuring a minimum quality of service. Fortunately, ML can be used to infer these interactions and provide a solution accordingly. In particular, through coordinated ML-assisted scheduling, different APs can trigger uplink/downlink transmissions from/to the appropriate stations (STAs), thus increasing the network throughput whilst reducing the number of packet collisions.

\subsubsection{Reinforcement Learning-Based Spatial Reuse} 
Spatial reuse aims to improve channel utilization through sensitivity adjustment mechanisms. However, selecting the best sensitivity threshold is not trivial given the complex spatial interactions that occur among devices. As a potential solution, reinforcement learning can be applied locally to improve spectral efficiency in a decentralized manner.

\section{ITU Unified Architecture for Future Networks}
\label{section:itu_architecture}

\begin{figure*}[!ht]
	\centering
	\includegraphics[width=0.55\textwidth]{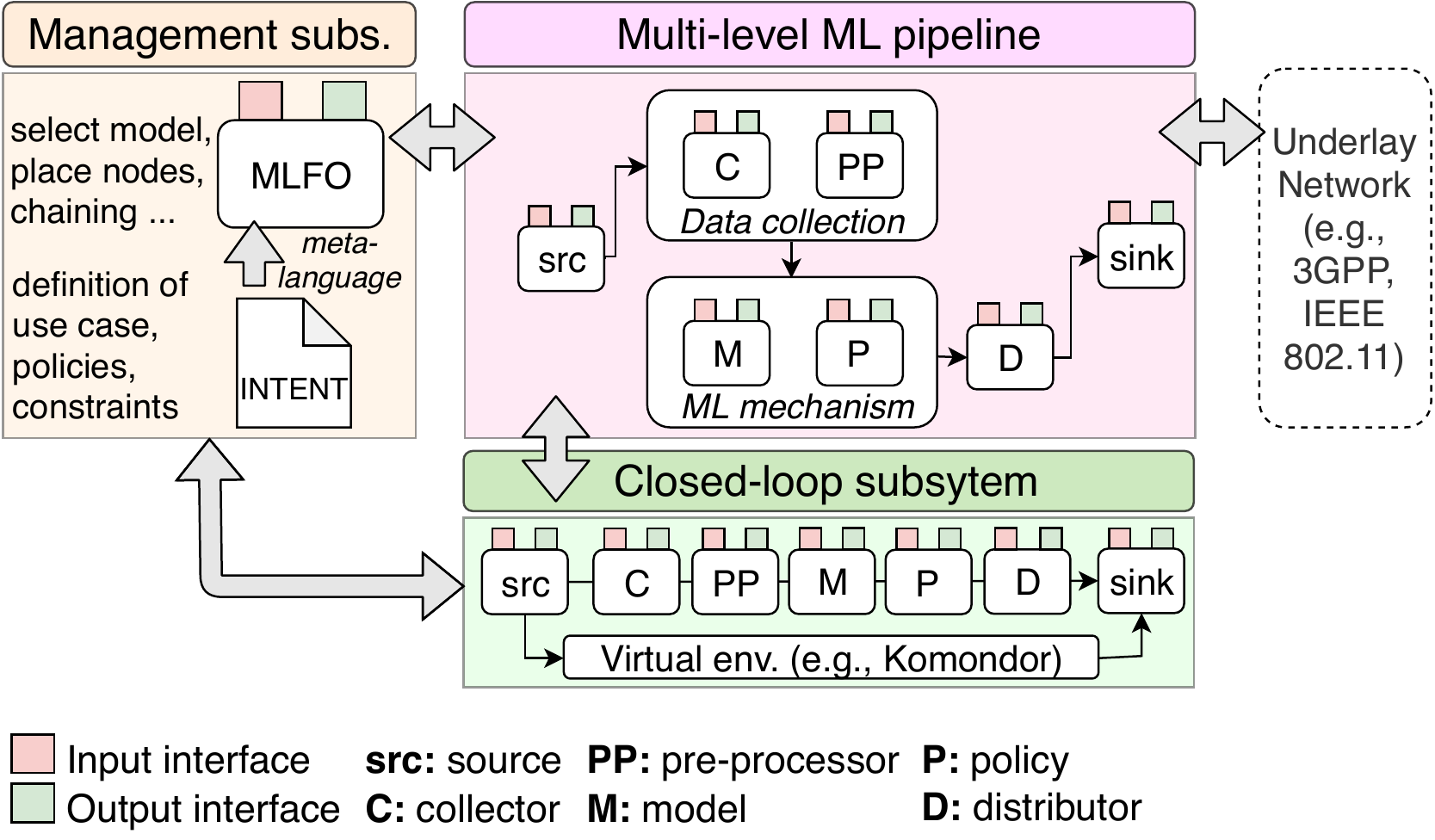}
	\caption{ITU's logical architecture for future networks \cite{itu2019architecture}. Entities contain input/output interfaces for communication, while the ML intent is a declarative file with information related to the use case.}
	\label{fig:itu_ml_architecture}
\end{figure*}

The FG-ML5G was created in November 2017 by its parent group, the ITU-T Study Group 13, to study the integration of ML mechanisms into future networks. This includes the definition of interfaces, protocols, data formats, and architectures. During its lifetime, the FG-ML5G has released several reports and contributions. Among them, we highlight the ITU's logical interoperable architecture for future networks \cite{itu2019architecture}, which defines an ML overlay that operates on the top of any unspecified underlay network technology (e.g., 3GPP, EdgeX, IEEE 802.11). The ITU's architecture aims to fulfill a set of technology-agnostic high-level requirements to support ML. For instance, the architecture must be able to support multiple types of data, thus taking advantage of heterogeneous data sources. 

Figure \ref{fig:itu_ml_architecture} shows the elements that compose the ML overlay (management subsystem, multi-level ML pipeline, and closed-loop subsystem). These elements are further described in the following subsections. Based on this standard overlay, ML applications can be instantiated in the logical entities (represented by white boxes).

\subsection{Management Subsystem} 
The management subsystem is in charge of the deployment and the orchestration of the ML services that operate in the underlying network. To that purpose, the Machine Learning Function Orchestrator (MLFO) entity is defined. The MLFO is first instantiated by a declarative intent that uses a meta language. It specifies the ML use case to be applied, including initialization, policies, and constraints. Then, the MLFO initializes the elements of the ML pipeline and monitors their operation during execution.

\subsection{Multi-Level Machine Learning Pipeline} 
The multi-level ML pipeline performs the actual ML operation in a given network underlay and it is in charge of the data collection, model application, and output distribution. The following logical entities compose the ML pipeline:
\begin{itemize}
	\item \textbf{Source (src):} generates data to be used by the ML mechanism.
	\item \textbf{Collector (C):} collects the data generated by sources.
	\item \textbf{Pre-processor (PP):} prepares the data collected for its utilization by the ML mechanism.
	\item \textbf{Model (M):} applies the ML model specified by the intent.
	\item \textbf{Policy (P):} provides a set of constraints and/or guidelines that delimit the behavior of the model.
	\item \textbf{Distributor (D):} spreads the ML output across all the corresponding targets (or sinks).
	\item \textbf{Sink (sink):} applies the ML output that is received from the distributor.
\end{itemize}

\subsection{Closed-loop Subsystem} 
In order to address network dynamics, the ML operation is assisted by a closed-loop subsystem, which can provide information to the system beforehand. As for the ML pipeline, the closed-loop subsystem is orchestrated by the management subsystem. In particular, a sandbox can be formed of real devices (pre-production internal network) or even be virtual (simulator/emulator). Network simulators such as ns-3 and Komondor \cite{barrachina2019komondor} are examples of closed-loop subsystems and can serve two purposes: \emph{i)} generate synthetic data for training, and \emph{ii)} run simulations to validate potential solutions before being applied in production.

\section{Machine Learning-Aware Architecture for IEEE 802.11 WLANs}
\label{section:wlans_architecture}

Based on their independence degree in terms of management and operation, WLAN deployments can be divided into two main families:
\begin{itemize}
	\item \textbf{Enterprise:} a set of BSSs can be jointly operated from the edge and/or the cloud, thus providing management and orchestration functionalities such as centralized authentication, or channel allocation. Enterprise-like deployments are realized through Extended Service Sets (ESS) and can be typically found in environments controlled by a single network operator, like university campuses, offices, stadiums, etc. 
	\item \textbf{Residential:} each BSS is responsible for its own management and operation. In the context of residential scenarios (but not limited to), peer-to-peer deployments are gaining popularity for infrastructureless communications (e.g., Wi-Fi direct).
\end{itemize}

\begin{figure}[ht!]
	\centering
	\includegraphics[width=\columnwidth]{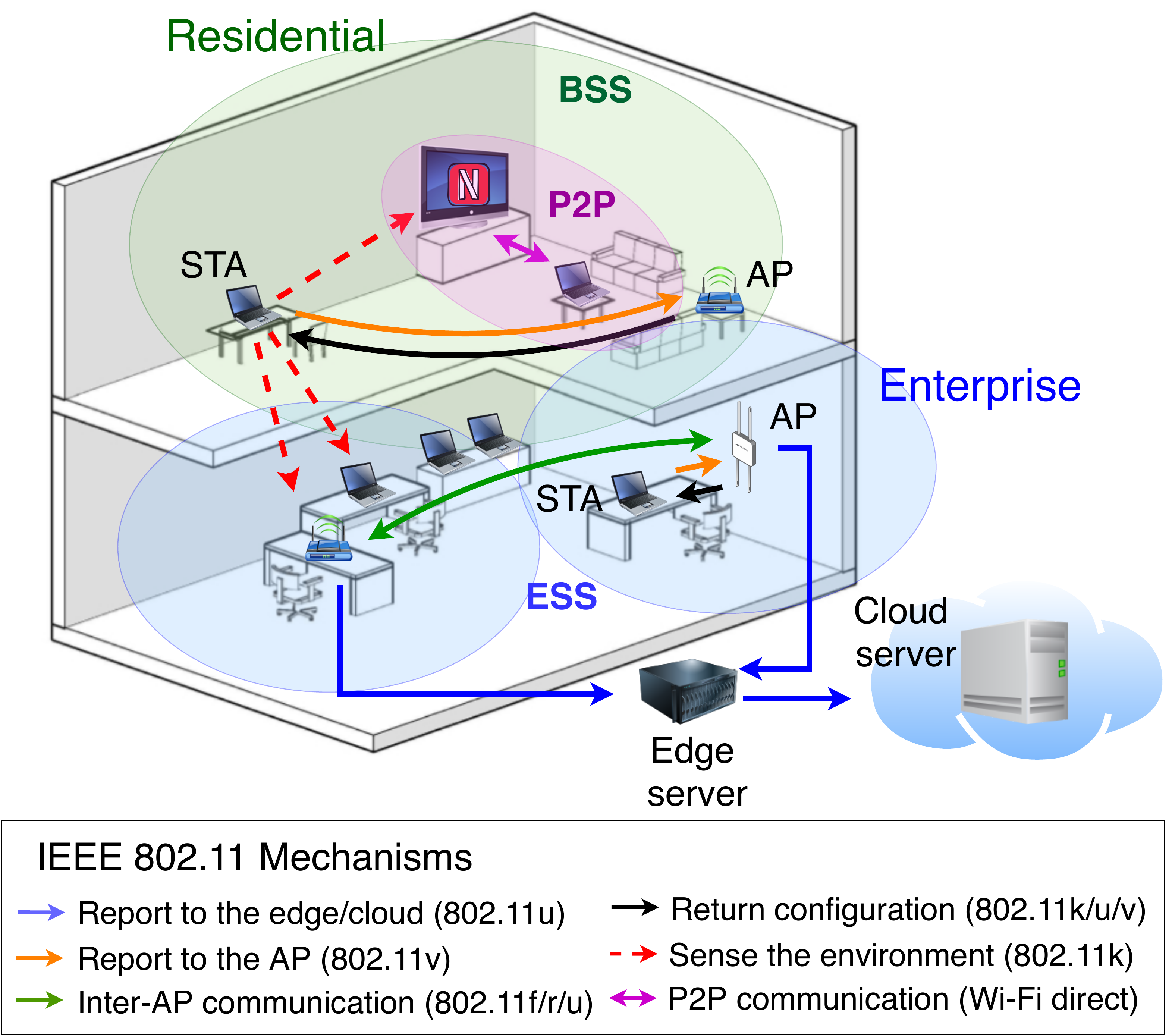}
	\caption{Enterprise and residential-like deployments and complementary IEEE 802.11 mechanisms to enable the utilization of ML.}
	\label{fig:overview_learning_approaches}
\end{figure}

Figure \ref{fig:overview_learning_approaches} illustrates the enterprise and residential-like deployments as well as a set of mechanisms that can facilitate the adoption of the ML-based architecture in WLANs. The following functionalities are provided:
\begin{itemize}
	\item \textbf{Information gathering (802.11k/r/v):} ML mechanisms can use information about the network topology and RF measurements to infer the behavior of other devices, or to extract important environmental characteristics.
	\item \textbf{Interoperability (802.11f/u):} Interoperability enables coordinated operations (e.g., scheduling, resource allocation), thus allowing to apply centralized/coordinated mechanisms such as in federated learning.
	\item \textbf{Security (802.11w):} ML mechanisms can use management frames that are protected so that a higher level of security is granted.
	\item \textbf{Validation (802.11t):} Performance evaluation in WLANs through test metrics can be useful to define optimization goals within the ML  operation.
\end{itemize}

\subsection{Challenges in IEEE 802.11 WLANs}
\label{section:ieee_80211_wlans}
The application of ML methods in WLANs is tightly tied to the technological challenges posed by these types of networks. The major challenges encountered in wireless communications stand for fast data expiry and lack of resources for data handling (e.g., storage, computation, and information exchange). Regarding Wi-Fi networks, we find the following challenges:
\begin{enumerate}
	\item \textbf{Non-stationarity:} channel fluctuations due to multipath fading, mobility of users and varying traffic needs entail a big challenge to ML applications. As a result of network dynamics, adaptability should be granted by continuously retraining ML models.
	\item \textbf{Limited communication resources:} since Wi-Fi operates under unlicensed bands, resources are scarce and shared. Hence, any potential communication required by a certain ML mechanism (as for distributed learning) may fail or be delayed if the medium is congested. As a result, the ML operation must be robust and resilient enough to react to potential communication issues.
	\item \textbf{Limited computation and storage resources:} computation and storage resources may also be scarce in WLANs, especially in residential-like deployments. Therefore, the ML operation should include computation-efficient procedures. Another implication of limited resources lies in the availability of information to be used by ML algorithms, especially for online learning methods.
	\item \textbf{Adversarial environment:} in many cases, Wi-Fi deployments are chaotic in the sense that many overlapping BSSs coexist without cooperation. This is a particularly interesting challenge for ML methods, where competition among agents may lead to an adversarial setting. Moreover, multi-vendor devices may implement different ML mechanisms, leading to clashing interests.
	\item \textbf{Legacy devices:} BSSs may coexist with other legacy devices that do not perform any intelligent operation. It is then required for ML methods to be aware of those devices, so that unfair situations are avoided.
\end{enumerate}

Apart from the previous WLAN-specific challenges, other inter-domain issues should be considered. For instance, end-to-end security is required since ML mechanisms store and/or exchange sensitive data that may be exposed. Besides, interoperability should be tackled when deploying ML solutions to different underlay networks. In this regard, the standardized ITU ML pipeline stands up as a promising solution.

\subsection{Computation Paradigms in IEEE 802.11 WLANs}
The various types of WLAN deployments and their computation and communication capabilities are closely linked to the type of ML solutions that can be applied to them: \emph{cloud} or \emph{edge-oriented}.

Cloud-oriented ML applications are characterized by
bearing high computational and storage resources, thus allowing them to collect various types of data from multiple sources, and to provide global and long-term solutions. The major challenge for cloud-oriented methods lies in the management of data and the corresponding synchronization, availability, and heterogeneity issues.

In edge-oriented mechanisms, the ML operation is mainly ruled by edge devices (e.g., APs and/or STAs), which, contrary to the cloud approach, typically lack powerful computation and storage resources. In consequence, edge-oriented mechanisms may only allow using simple and lightweight computing ML algorithms. Nevertheless, edge servers can be added to deploy more powerful solutions promptly. The edge-oriented approach is useful for real-time ML applications that manage local (and even highly-varying) information.

Apart from cloud and edge-oriented settings, we may distinguish between methods based on their cooperation degree. In cooperative approaches, nodes interact among them for the sake of jointly conducting the learning operation (e.g., sharing a reward). However, reliable and timely connections among learners are often required. In this regard, \cite{lin2017deep} showed the role of communications on speeding up a distributed training procedure over a set of nodes in a network. Alternatively, for the non-cooperative case, the learning operation may lead to adversarial settings, especially since BSSs share resources such as the spectrum.

\subsection{Realization of the ML-Aware Architecture for WLANs}

\begin{figure}[ht!]
	\includegraphics[width=1\columnwidth]{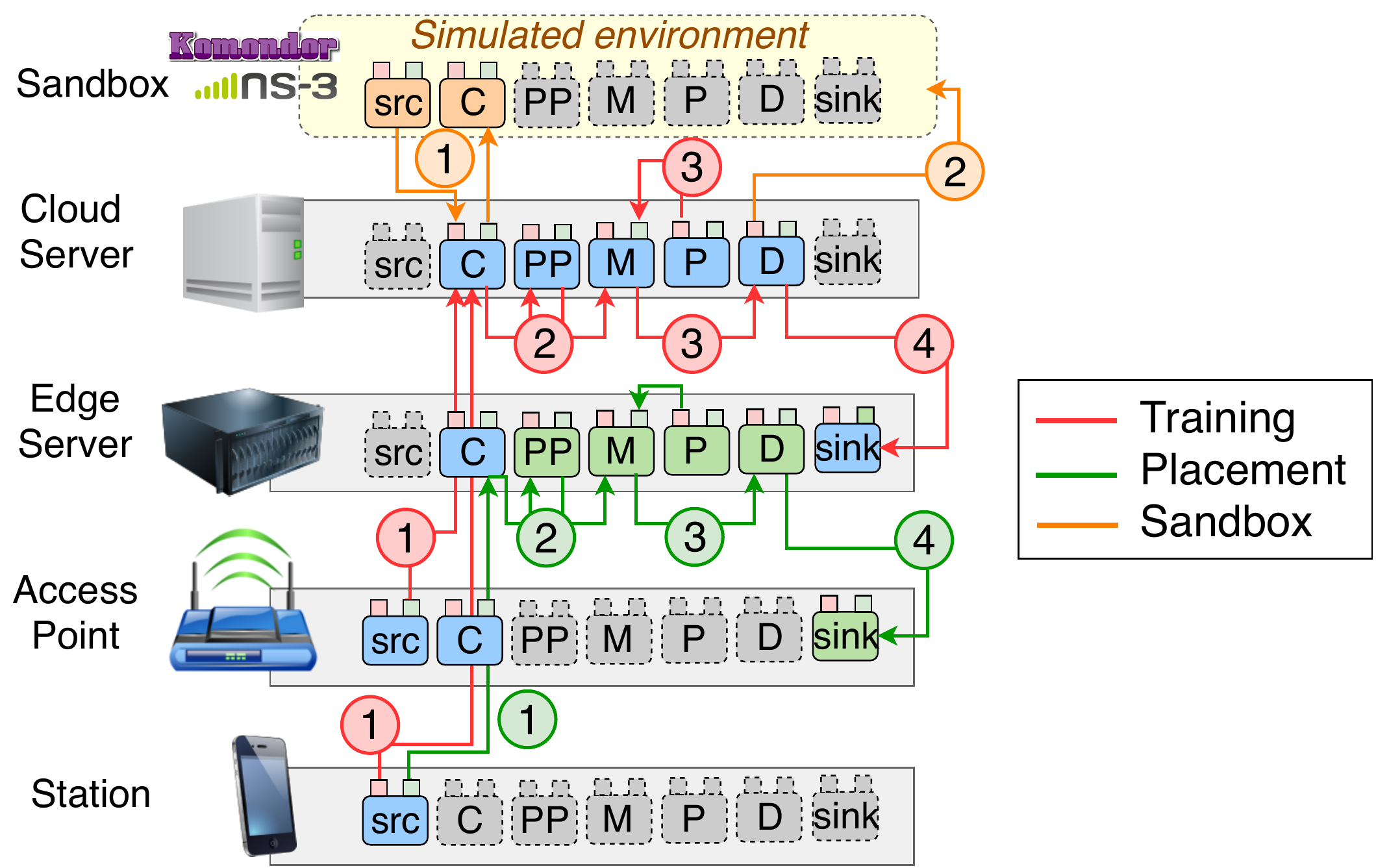}
	\caption{Realization of the ITU's ML architecture for IEEE 802.11 WLANs through a hybrid ML-based solution for AP (re)association and handover.}
	\label{fig:ml_architecture_wlan}
\end{figure}

To showcase the adoption of the architecture, let us retake the AP (re)association and handover example (see Fig. \ref{fig:ml_architecture_wlan}). We now consider a hybrid solution where two main ML-based processes are held: training (learn from data) and placement (apply the learned knowledge). 

While the training procedure is carried out at the cloud (collect data from multiple sources), the placement operation is done at the edge (provide timely responses to new cases). Notice that the system can also be re-trained during the placement phase, based on newly acquired local data.

Specifically, the training procedure consists of the following steps (shown in red):
\begin{enumerate}
	\item \textbf{Data collection:} the cloud server collects information of different kinds from APs and STAs, such as user information (e.g., location), performance (e.g., delay), application data (e.g., traffic load), or channel status reports (e.g., sensed interference). This information can be used either for training or feeding auxiliary algorithms that help the main AP association procedure (e.g., predict user behavior).
	\item \textbf{Pre-processing:} the data collected at the cloud is pre-processed so that the ML method can properly manage it. For instance, in case of applying a multiple linear regression, the input information needs to be converted into normalized features (i.e., convert the rate given in Mbps into a scalar between 0 and 1).
	\item \textbf{Model generation:} when generating the ML model, certain policies need to be considered. For instance, an AP may set a maximum number of associated STAs. The policies are strongly tied to the capabilities of the devices or the existing regulations (e.g., maximum regulated transmission power).
	\item \textbf{Output distribution:} once the ML method in the cloud generates the output (i.e., the predicted function for new (re)associations), it is distributed throughout the sink edge servers, which are then ready to give quick response to new cases.
\end{enumerate}

In the placement phase (shown in green), we find:
\begin{enumerate}
	\item \textbf{Handle new requests:} new (re)association requests or potential handovers are detected based on newly acquired information from STAs. This information is collected by the edge server.
	\item \textbf{Pre-processing:} the acquired information is then processed by the edge server, just like for the training phase.
	\item \textbf{Run the ML solution:} the ML method provided by the cloud is applied locally at the edge server, which provides an output for the new request.
	\item \textbf{Apply the ML solution:} the (re)association decision is distributed to the corresponding AP.
\end{enumerate}

Finally, it is worth pointing out the role of the sandbox, which can be mainly twofold (shown in orange):
\begin{enumerate}
	\item \textbf{Generate data for training:} the sandbox can act as a source in the ML pipeline by generating synthetic data for training purposes. Nevertheless, the data provided by the sandbox is limited to several factors such as the accuracy of the simulation model or the degree of similarity between the sandbox and the real network. 
	\item \textbf{Preliminary model testing:} alternatively, the sandbox can be used to validate the output of the ML method before being applied to the real network.
\end{enumerate}

\begin{figure}[ht!]
	\centering
	\includegraphics[width=\columnwidth]{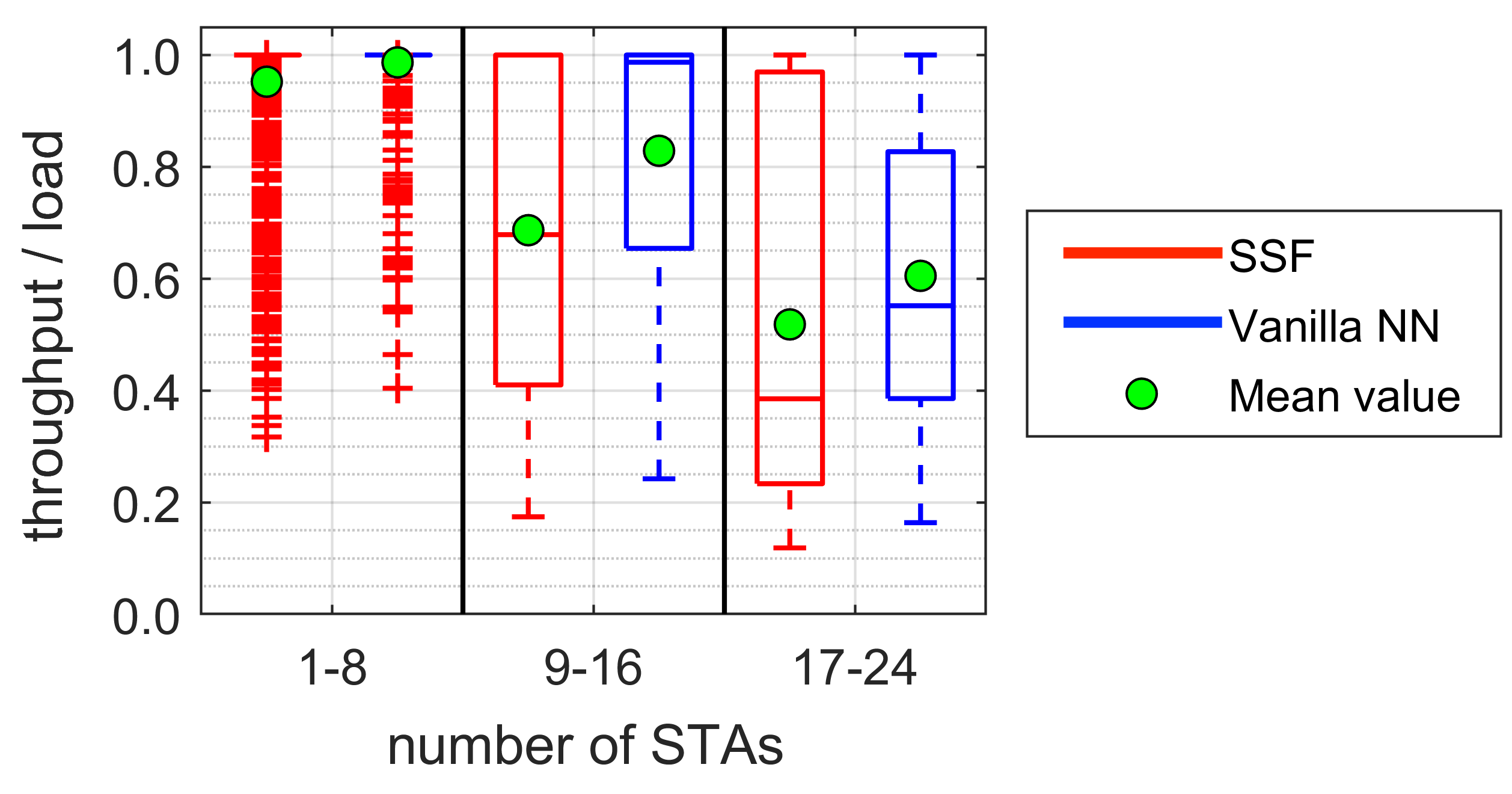}
	\caption{Performance evaluation of the AP association problem in WLANs: SSF versus Neural Network (NN). The mean performance of each mechanism is represented by a green dot.}
	\label{fig:results_use_case}
\end{figure}

To showcase the potential of the ML-based architecture through numerical results,\footnote{Given the novelty of the technologies studied in this paper, our results have been obtained from well-know standard-compliant models, hence their accuracy is tied to them. Nevertheless, this is a first step to understand the potential benefits of using an ML-based architecture in next-generation wireless networks. For the sake of reproducibility and disclosure, all the source code is open and publicly available at \href{https://github.com/fwilhelmi/machine_learning_aware_architecture_wlans}{https://github.com/fwilhelmi/machine\_learning\_aware\_architecture\_wlans}, accessed on Jan. 31, 2020.} we compare the performance of classical AP association procedure (SSF) against a novel ML-based approach (based on vanilla neural networks). In particular, the neural network predicts the throughput that an STA will obtain after associating to a given AP based on a set of features or characteristics (e.g., current load, received signal strength). Figure \ref{fig:results_use_case} shows the throughput received by each STA versus the load it generates, for different deployment densities. We observe that the ML approach improves the average performance and balances the results obtained by all the STAs. This is because the ML function can capture complex patterns from dense deployments, thus guaranteeing minimum throughput requirements to STAs (at the expense of missing the maximum performance peaks).

\section{Concluding Remarks}
Current networks are not yet prepared for the pervasive adoption of ML-based operation. Hence, disruptive architectural changes are required. For the sake of moving forward in this field, this article introduced the ITU's unified architecture for future networks and provided a realization for IEEE 802.11 WLANs. The different forms of Wi-Fi networks allow uplifting the flexibility characteristic of the ITU's architecture, thus enabling from edge to cloud-oriented solutions, including hybrid approaches. 

To conclude, future wireless networks are envisioned to share a common flexible architecture that allows a fast adaptation of resources to accommodate a plethora of ML-enabled verticals. Nevertheless, a lot of effort is still required before reaching fully intelligent wireless networks. Among several open issues, we highlight the ones related to data handling (\textit{how/where to store data? how to assess the expiry of data?}), orchestration (\textit{how to distribute the ML operation? how to deal with heterogeneity?}), and robustness of the ML methods (\textit{how to deal with uncertainty? how to prevent unprecedented events}?). 

\section*{Acknowledgment}

This work has been partially supported by grants MDM-2015-0502, 2017-SGR-11888, by WINDMAL PGC2018-099959-B-I00 (MCIU/AEI/FEDER,UE), by a Gift from the Cisco University Research Program (CG\#890107) Fund, and by SPOTS project (RTI2018-095438-A-I00) funded by the Spanish Ministry of Science, Innovation and Universities. The work by Sergio Barrachina-Mu\~noz is supported by an FI grant from Generalitat de Catalunya.

\ifCLASSOPTIONcaptionsoff
  \newpage
\fi

\bibliographystyle{IEEEtran}
\bibliography{bibliography}

\begin{IEEEbiographynophoto}{Francesc Wilhelmi}
(francisco.wilhelmi@upf.edu) holds a B.Sc. degree in Telematics Engineering (2015) and an M.Sc. in Intelligent and Interactive Systems (2016), both from Universitat Pompeu Fabra (UPF). He is currently pursuing a Ph.D. in Information and Communication Technologies at UPF.
\end{IEEEbiographynophoto}

\begin{IEEEbiographynophoto}{Sergio Barrachina-Mu\~noz}
(sergio.barrachina@upf.edu) obtained his B.Sc. degree in Telematics Engineering and his M.Sc. in Intelligent Interactive Systems in 2015 and 2016, respectively, both from Universitat Pompeu Fabra (UPF), Barcelona. Currently, he is a PhD student and teacher assistant in the Wireless Networking research group at UPF.
\end{IEEEbiographynophoto}

\begin{IEEEbiographynophoto}{Boris Bellalta}
(boris.bellalta@upf.edu) is an Associate Professor in the Department of Information and Communication Technologies (DTIC) at Universitat Pompeu Fabra (UPF). He is the head of the Wireless Networking research group at DTIC/UPF.
\end{IEEEbiographynophoto}

\begin{IEEEbiographynophoto}{Cristina Cano}
(ccanobs@uoc.edu) holds a Ph.D. (2011) in Information, Communication and Audiovisual Media Technologies from Universitat Pompeu Fabra (UPF). She has been a research fellow in the Hamilton Institute of the National University of Ireland, Maynooth (2012-2014), in Trinity College Dublin (2015-2016) and in Inria-Lille in France (first half of 2016). Currently, she is an associate professor at Universitat Oberta de Catalunya (UOC). 
\end{IEEEbiographynophoto}

\begin{IEEEbiographynophoto}{Anders Jonsson}
(anders.jonsson@upf.edu) is the director of the Artificial Intelligence and Machine Learning group at Universitat Pompeu Fabra (UPF). He received his Ph.D. in computer science in 2005 from the University of Massachusetts Amherst, USA, and has been at UPF ever since.
\end{IEEEbiographynophoto}

\begin{IEEEbiographynophoto}{Vishnu Ram}
(vishnu.n@ieee.org) worked for Motorola/Nokia/Siemens in advanced technologies teams for 21 years. He was a Scientific Advisory Board Associate (SABA) member of Motorola Networks. He has published several drafts in IETF, contributed to ETSI, 3GPP in his role as a senior specialist (Radio Resource Management). He is currently working as an independent researcher.	
\end{IEEEbiographynophoto}

\vfill

\end{document}